\documentstyle[epsfig]{aipproc}

\begin{document}
\begin{flushright}
 May 21  1997\\
TAUP 2425/97
\end{flushright}

\title{A violation of the factorization theorem  for ``hard" inclusive
production}
\author{ Eugene   Levin }
\address{  School of Physics and Astronomy\\
 Raymond and Beverly Sackler Faculty of Exact Science\\
 Tel Aviv University, Tel Aviv, 69978, ISRAEL\\
and\\
 Theory Department, Petersburg Nuclear Physics
Institute\\
 188350, Gatchina, St. Petersburg, RUSSIA}

\maketitle

\begin{abstract}
This talk is a digest of Ref.\cite{GLMNM}, where 
 a new mechanism for hard inclusive production, which leads to a violation
of the factorization theorem (FT), is suggested.
 Numerical  
estimates for the effect  are  given for  high
energy hadron (nucleus)  scattering.
\end{abstract}
\section*{Introduction}
 In this talk  we consider a new mechanism for hard inclusive
production, which violates the factorization theorem \cite{FACT}.
 The suggested mechanism is
present in
 any central hard inclusive  production, such as heavy Higgs meson
production and/or the high $p_{T}$
inclusive
production of mini jets and jets.
We illustrate this mechanism for the case of inclusive heavy Higgs
production
in nucleon - deutron high energy scattering. 

 The usual description of inclusive production of a Higgs meson with mass
$M$
in a nucleon - deutron interaction, is based  on the factorization
theorem \cite{FACT},
using which the cross section of interest
is given  by
\begin{equation} \label{FT}
\sigma(Higgs)\,\,=\,\,\int  d x_1 d x_2 F^p_N (x_1,M^2) F^p_D (x_2,M^2)
\,\sigma(hard)\,\,,
\end{equation}
where $F^p_D(x_2,M^2)$ denotes the parton distribution within  the
deutron,
 and in the impulse approximation $ F^{p}_{D} = 2F^{p}_{N}$.
$\sigma(hard)$ denotes the cross section for  Higgs meson production in
the
parton - parton collision.

\section*{ Factorization theorem  and  AGK cutting rules.}

The factorization theorem (FT) can be proven for the nucleon  - deutron
scattering using the AGK cutting rules \cite{AGK}( see Ref. \cite{BARY}
 and references therein for
detail  discussion of the AGK cutting rules in QCD). Indeed, let us assume
that the FT holds for nucleon - nucleon scattering. For nucleon - deutron
scattering we have two contributions to the total cross section:
the projectile nucleon interacts with one nucleon in the deutron ( impulse
approximation ) and it interacts with both nucleons in the deutron (
Glauber corrections). The first process obviously leads to the FT of
Eq.(1), while for the second one we can use the AGK cutting rules for the
decomposition of the total cross section with respect to the multiplicity
of
produced particles. The relation between different contributions is:
\begin{equation} \label{AGK}
\sigma^{(0)}_{tot}\,:\,\sigma^{(1)}_{tot}\,:\,\sigma^{(2)}_{tot}\,\,=\,\,(
1\,+\,\rho^2)\,:\,-\,4\,:\,2\,\,,
\end{equation}
where $\rho\,=\,\frac{Re A_N}{Im A_N}$ and $A_N$ is the amplitude of the
nucleon - nucleon interaction. $\sigma^{(0)}_{tot}$ is the cross section
of all processes where no particles were produced in the central rapidity
region ( elastic scattering and single and double diffractive
dissociation); $\sigma^{(1)}_{tot}$ is the contribution to the total
inelastic cross section, which has the same multiplicity distribution as
in nucleon - nucleon inelastic cross section, and $\sigma^{(2)}_{tot}$ is
part of the total inelastic cross section for the nucleon - deutron
scattering which describes the inelastic interaction of the projectile
nucleon with two nucleons in the deutron.

 One can see that Eq.(2) leads to
the cancellation of the interaction with the two nucleon in the deutron
for the inclusive cross section of the particle in the central rapidity
region. Indeed, $\sigma^{(0)}_{tot}$ does not contribute to this
production while two other contributions cancels since in the subprocess
described by $\sigma^{(2)}_{tot}$ there is two ways of the production of
the particle of the interest. Therefore, the inclusive production can be
described in the impulse  approximation for which the FT is valid by our
assumption. The natural question arises: {\it what is wrong in our above
discussion}.
 
\section*{The AGK cutting rules and the interference diagrams.}

Let us start with the short answer to the formulated question: {\it we
missed the interference diagrams given in Fig.1} and these diagrams 
give the new mechanism for centrally produced Higgs meson ( or  for any
other ``hard" process.
\begin{figure}
\centerline{\psfig{file=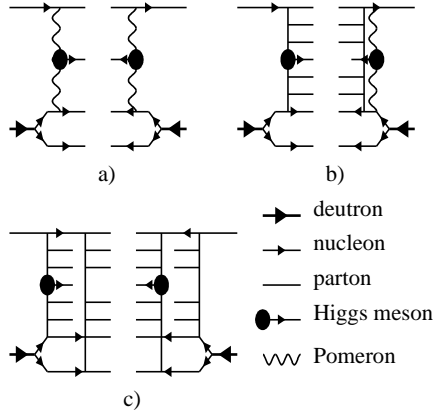,width=80mm}}
\caption{Interference diagrams for different inclusive contributions:
(a)$\sigma^{(0)}_2$, (b) $\sigma^{(1)}_2$ and (c) $\sigma^{(2)}_2$.}
\end{figure}

 Trying to describe what kind of processes we picture in Fig.1,
one can see that Fig.1a shows the process where the
Higgs, which is  produced  centrally, is separated
 by two large rapidity gaps (LRG)
  from the small final state multiplicities
which  occur on the edges of the rapidity plot.
We define this process as a double Pomeron exchange reaction, and denote
its
contribution as   $\sigma^{(0)}_2$. The contribution  of Fig.1b is denoted
by $\sigma^{(1)}_2$. This is a mixed diagram, where the Higgs which is
produced in a double Pomeron process, is superimposed  on a normal
uniform rapidity distribution typical of an inelastic nucleon-nucleon
reaction. Finally, Fig.1c describes Higgs production as part
of the nucleon-nucleon background to the rapidity distribution. We denote
this contribution by $\sigma^{(2)}_2$. To obtain our final result we need
to
sum over all three of the above contributions, noting that these
  are not necessarily positive.

 \par The nucleon-nucleon (NN) amplitude for Higgs production via  
 double Pomeron exchange, has been calculated \cite{BL}, to be
\begin{equation}  \label{HP}
A_H\,=\,A(NN\,\rightarrow\,N + (LRG) + H + (LRG) + N)\,\,=\,\,2 g_H
A_P\,\,,
\end{equation}
 $g_H$ is the the vertex of the hard parton - parton $\rightarrow $ Higgs
process and  $A_P$ is the Pomeron exchange amplitude.

The second ingredient in our calculation is the amplitude shown in Fig.1c 
( see also one of the amplitude in Fig.1b).
This amplitude has no analog in the case of a single nucleon-nucleon
 interaction,  
and it depicts the cut in the diagram shown in  Fig.1a. We note that
this diagram is equal to $Im A_H$, unlike the case for inelastic
 nucleon-nucleon cross section where  $\sigma_{in} = 2 Im A_P$. The above
result follows  from the unitarity constraint \cite{AGK}.

\par Recalling that
 the  integration over the longitudinal components of the momentum
carried by the Pomeron
in
Fig.1b results in  a  negative sign for the interference diagram (see Ref.
\cite{AGK} for details),  we can easily calculate
all diagrams contributing to the inclusive Higgs meson production.
Indeed, calculating $\sigma^{(0)}_2$,  the sum of  the diagrams in Fig.1a,
and 
using Eq.(3)
, one has

\begin{equation}
\sigma^{(0)}_2\,\,=\,\,4 g^2_H \,\{\, (ReA_P)^2\,\,+\,\,(ImA_P)^2\,\}\,\,.
\end{equation}
For the sum of diagrams in Fig.1b  we have
\begin{equation}
\sigma^{(1)}_2\,\,=\,\,-\,8\,g^2_H\,(Im A_P)^2\,\,.
\end{equation}
Finally, for $\sigma^{(2)}_2$, shown in Fig.1c, we obtain
\begin{equation}
\sigma^{(2)}_2\,\,=\,\,4\,g^2_H\,( Im A_p)^2\,\,.
\end{equation}
The relation between the  different  contributions
is  
\begin{equation}
\sigma^{(0)}_2\,:\,\sigma^{(1)}_2\,:\,\sigma^{(2)}_2\,\,=\,\,(1+
\rho^2)\,:\,-\,2\,:\,1\,\,,
\end{equation}
where $\rho = \frac{Re A_P}{Im A_P}$. 

Summing all contributions we obtain an additional term in the cross
section compare to the one
given in Eq.(1)
\begin{equation}
\sigma(N  + D \,\rightarrow\,H + X)\,\,=\,\,4\,g^2_H\,(Re A_P)^2\,\,.
\end{equation}

We would   like to draw  attention to the novel fact,
 that the contribution of the new
mechanism  is proportional to the real part of the amplitude.
For a soft Pomeron with $\alpha_{P}(0)\; =\; 1$, the contribution of
Eq.(8)
vanishes, and we recover the factorization theorem \cite{FACT}, i.e.
an exact cancellation of the diagrams shown in Fig.1c. However, for the
hard processes ( ``hard" Pomeron) both theory and experiment lead to a
steep behaviour of the cross section as a function of energy ( see Ref.
\cite{GLMNM}) which results in considerable violation of the factorization
theorem for totaly inclusive production. Indeed, at high energy ( low $x$)
$\rho$ is equal to
\begin{equation}
\rho\,\,=\,\,\frac{Re A_P}{Im A_P} = \frac{\pi}{2}\,\frac{\frac{d Im
A_P}{d \ln(1/x))}}{ Im A_P}\,\,=\,\,<\omega>\,\,.
\end{equation}
In all available parametrization for the deep inelastic structure function
$<\omega> \,\,\approx\,\,0.3 - 0.5$ at $ 10^{-2} \,< \,x\,<\,10^{-5}$.

\section*{Numerics.}
To give a quantative measure of the effect of the nonfactorization
  we plot in Fig.2 the ratio of the nonfactorized contribution ( see
Ref.\cite{GLMNM} for the formula of this contribution)
 to the cross section from the factorization theorem of Eq.(1), namely
\begin{equation} 
R\,\,=\,\,\frac{2 \pi^2}{R^2_1 R^2_2}\frac{
 | \int^{\frac{M^2}{4}}_{Q^2_0} d k^2_t \frac{\partial \phi_1(x_1,k^2)}
{\partial \ln (1 /x_1)}\,\phi_2(x_2,k^2)|^2}{ x_1 G_1(x_1,\frac{M^2}{4})
x_2 G_2(x_2,\frac{M^2}{4})}\,\,,
\end{equation}
where $x_i G_i (x_i,M^2)$ is the gluon density  of the $i$-th
hadron and $\phi$ is the unintegrated gluon density, which closely related
to the gluon structure function,namely,
$\alpha_S(Q^2) x G(x,Q^2)\,=\,\int^{Q^2}\,\,d k^2_t
\alpha_S(k^2)\phi(x,k^2)$. 
In Fig.2 we took $y$ = 0, where $y$ is rapidity of produced Higgs meson.
 Notice, that this ratio does not depend
on the hard cross section, and as such is also applicable to any   
central
production process, in particular, hard minijet and jet production.
In evaluating  
Eq.(10)
we use  \cite{GLMNM} 
the GRV parametrization
\cite{GRV}
 for the gluon density.
We assume that  $R^2_1 =
R^2_2  = 5\,GeV^{-2}$ (see Ref.\cite{GLMSLOPE} ).
For the initial virtuality we take $Q^2_0 = 1\, GeV^2$, as  the GRV
 parametrization 
is in agreement with
  the HERA data
on $F_2(x,Q^2)$ \cite{HERA}
 for all $Q^2
\,\geq\,1\,GeV^2$.
\begin{figure}[hbt]
\centerline{\psfig{file=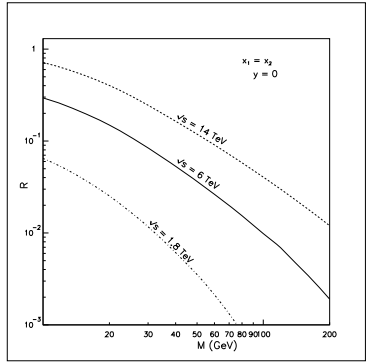,height=60mm}}
\caption{Calculated R (see Eq.(10))}
\end{figure}

One can see, that $R$ is quite big for low masses and decreases
when M increases.  For example, $R$ = 1.2   for $M = 10\; GeV$.
 We do not expect
the  Higgs meson  mass to be that small, but our result is also
applicable in
the case of
  jet production   where   $M \approx 2
p_{\perp}$.  Accordingly, with  such a value of $R$ we can expect  minijet
production (jets with $p_{\perp} \approx 5 GeV$ ) which can be responsible
for the structure of the minimum bias events at the Tevatron energy.

The value of $R$ is bigger for  nucleus - nucleus interaction.
 To estimate
this value we need to multiply Eq.(10) by a factor of  $A^{\frac{1}{3}}_{1
eff}
\,A^{\frac{1}{3}}_{2 eff}$. Using the simple relation
$R^2_A\,=\,r^1_0 A^{\frac{2}{3}}$, one has
$A^{\frac{1}{3}}_{eff}\,=\,\frac{R^2_1}{r^2_0} (\,A^{\frac{1}{3}}
\,-\,1\,)
\,+\,1$. Therefore, for a  gold - gold interaction we  expect that
 $R$ is enhanced  by factor 4 - 9.

\section*{Conclusions.}
Our conclusion is very simple: {\it the factorization theorem \cite{FACT}
is only approximate} and the violation of the FT can be rather big
especially for heavy ion collisions and/or for hadron ones at the LHC
energies. 
We firmly believe that the formal proof of the FT given in Ref.\cite{FACT}
should be reconsidered to find out the correct region of the applicability
of this theorem.

\end{document}